# A penalty criterion for score forecasting in soccer

Jean-Louis Foulley (1), Gilles Celeux (2)

1. Institut Montpellierain Alexander Grothendieck, France, foulleyjl@gmail.com
2. INRIA-Select, Institut de mathématiques d'Orsay, France, gilles.celeux@math.u-psud.fr

## Abstract

This note proposes a penalty criterion for assessing correct score forecasting in a soccer match. The penalty is based on hierarchical priorities for such a forecast i.e., i) Win, Draw and Loss exact prediction and ii) normalized Euclidian distance between actual and forecast scores. The procedure is illustrated on typical scores, and different alternatives on the penalty components are discussed.

***Keywords:*** soccer competition; quantitative analysis of sports; exact score forecasting; distance and metrics.

## Introduction

Methods of forecasting outcomes of soccer matches are usually based on the rate of correctly predicted results in terms of win (W), draw (D) and loss (L) or on scoring rules for probability forecasts of each of these three outcomes via e.g. the Brier or the logarithm score (Groll et al., 2015; Gneiting & Raftery,2007; Foulley, 2015).

Assessing the goodness of forecast scores is a challenging task especially in soccer as illustrated by the following example of an actual score (2, 1). Using a square error (SE) criterion, forecast (0, 1) gives SE = 4 while (0, 0) gives SE = 5 which is counterintuitive as long as a draw (0, 0) is usually preferred to a loss forecast (0, 1) for a win (2, 1).

Therefore, closeness between actual and forecast numbers of goals scored does not completely retrieve what benefit a team can expect at the end of the game which is first a win, at least a draw, and secondly, if possible in the case of a win, with the largest goal difference. This is reflected by the UEFA & FIFA ranking of teams with 3, 1 and 0 points for a win (W), draw (D) and loss (L) respectively followed by the largest number of goal difference. Therefore, there is a clear hierarchy in the objectives which should be considered in defining the accuracy of forecasting with the additional difficulty that the set of outcomes is on an ordinal scale W>D>L rather than a nominal scale (Constantinou & Fenton, 2012). The objective of this note is to propose a criterion measuring the forecasting errors in a proper way. The first part of this note is devoted to the method proposed to address this issue followed by a numerical illustration. Finally results and methods are discussed.

## The penalty forecasting criterion

To meet the above-mentioned constraints, a solution consists of setting a penalty $C$ pertaining to the categories of Actual (A) and Forecast(F) scores and add an extra term within each category to take into account the variability of scores therein.

Letting the actual and forecast goals scored in a match be noted $\mathbf{g}^A = (g_1^A, g_2^A)$, and $\mathbf{g}^F = (g_1^F, g_2^F)$ respectively, we suggest considering the following expression of a forecasting penalty (FP)

$$FP\left(\mathbf{g}^A, \mathbf{g}^F\right) = C\left(\mathbf{g}^A, \mathbf{g}^F\right) + D\left(\mathbf{g}^A, \mathbf{g}^F\right) \tag{1}$$

where

$C = 0$ if actual $\mathbf{g}^A$ and forecast $\mathbf{g}^F$ are in the same category (W or D or L),

$C = c_0$ if actual and forecast are in adjacent categories (W,D) or (D,L), and

$C = 2c_0$ if actual and forecast are in opposite categories (W,L) or (L,W),

$c_0$, being a positive scalar the value of which is discussed later on.

The second term is chosen as the normalized Euclidian $L_2$-distance

$$D\left(\mathbf{g}^A, \mathbf{g}^F\right) = \frac{\left\|\tilde{\mathbf{g}}^A - \tilde{\mathbf{g}}^F\right\|_2}{\left\|\tilde{\mathbf{g}}^A\right\|_2 + \left\|\tilde{\mathbf{g}}^F\right\|_2}, \tag{2}$$

where $\|\mathbf{u}\|_2 = \left(\sum_{i=1}^{2} u_i^2\right)^{1/2}$ and $\tilde{\mathbf{g}}^I = \left[f\left(g_1^I\right), f\left(g_2^I\right)\right]$ for $I = A, F$ is a monotonic transformation of the numbers of goals.

The Euclidian $L_2$ normalized distance provides a metric distance satisfying the triangular inequality $\|AF\|_2 \leq \|OA\|_2 + \|OF\|_2$ (Deza & Deza, 2009). Therefore, from a geometric point of view, $D$ being the length ratio $\|AF\|_2 / \left(\|OA\|_2 + \|OF\|_2\right)$, it lies between 0 and 1, so that $FP$ varies between 0 and $2c_0 + 1$. The value of $c_0$ depends on the amount of overlapping of $FP_2$ values between categories, one wants to accept. Mathematically, there is no overlapping whatever scores are, if $c_0 \geq 1$.

As the number of goals scored in a match can be modelled as Poisson variables (Dixon & Coles, 1997), we propose to work on a typical transformation of such random variables as for instance, the Anscombe transformation: $\tilde{g} = 2\sqrt{g + 3/8}$ , or the Freeman-Tukey transformation: $\tilde{g} = \sqrt{g} + \sqrt{g + 1}$. Such a transformation has two advantages: i) it stabilizes the variance of the distribution of goals and ii) it avoids degenerate penalty values when the A or F score lies in the origin such as $FP_r\left[(0,0),(x,x)\right] = 1$ for any $x > 0$.

The penalty rule defined in (2) for a single match can be averaged over a set of $n$ matches as those arising in a competition (for instance, $n$=64, 125 for the FIFA World Cup and UEFA Champions League respectively) to measure the mean forecasting penalty (MFP) pertaining to football scores guessed by a given individual or group:

$$MFP = \frac{1}{n}\sum_{j=1}^{n}\left(C_j + D_j\right). \tag{3}$$

**Numerical illustration**

The penalty rule is illustrated in Table 1 for typical numbers of goals scored. Results are listed according to the different doublets of Win, Draw and Loss results i.e., (D, D), (W, W), (W, D) and (W, L) categories with $c_0 = 1$.

Using the Anscombe transformation in the (D, D) avoids that the penalty associated to (0, 0) vs. (x, x) scores are all equal to 1 whatever the value of x and gives values increasing with x.

In the (W, W) category notice that forecast (3, 2) and (3, 1) are preferred to (1, 0) if actual score is (2, 1). For the same actual score (2, 1), the draws (1, 1) and (2, 2) give low values of D but, due to the addition of C, are dominated by Win forecast scores. In the (W, L) category, forecast (1, 2) and (2, 3) are better than (0, 1) when actual score is (2, 1).

Implementing C in the expression of FP is again well illustrated by the case of an actual score (1,0); forecast (0, 1) gives D = 0.285 while D = 0.387 for (0, 0). In contrast, the FP values 1.387 and 2.285 are in keeping with what one would expect.

Table 1: Penalty values for typical score values

| Actual | | Forecast | | Distance | | WDL | Penalty | |
|---|---|---|---|---|---|---|---|---|
| GA1 | GA2 | GF1 | GF2 | D1 | D2 | C | FP1 | FP2 |
| 0 | 0 | 1 | 1 | 0.314 | 0.314 | 0 | 0.314 | 0.314 |
| 0 | 0 | 2 | 2 | 0.431 | 0.431 | 0 | 0.431 | 0.431 |
| 0 | 0 | 3 | 3 | 0.500 | 0.500 | 0 | 0.500 | 0.500 |
| 1 | 1 | 2 | 2 | 0.136 | 0.136 | 0 | 0.136 | 0.136 |
| 2 | 1 | 1 | 0 | 0.206 | 0.206 | 0 | 0.206 | 0.206 |
| 2 | 1 | 3 | 2 | 0.109 | 0.109 | 0 | 0.109 | 0.109 |
| 2 | 1 | 4 | 3 | 0.183 | 0.183 | 0 | 0.183 | 0.183 |
| 2 | 1 | 3 | 1 | 0.052 | 0.072 | 0 | 0.072 | 0.072 |
| 2 | 1 | 4 | 2 | 0.145 | 0.146 | 0 | 0.146 | 0.146 |
| 2 | 1 | 0 | 0 | 0.378 | 0.387 | 1 | 1.378 | 1.387 |
| 2 | 1 | 1 | 1 | 0.073 | 0.103 | 1 | 1.073 | 1.103 |
| 2 | 1 | 2 | 2 | 0.064 | 0.090 | 1 | 1.064 | 1.090 |
| 2 | 1 | 0 | 1 | 0.206 | 0.285 | 2 | 2.206 | 2.285 |
| 2 | 1 | 1 | 2 | 0.136 | 0.135 | 2 | 2.136 | 2.135 |
| 2 | 1 | 2 | 3 | 0.109 | 0.153 | 2 | 2.109 | 2.153 |
| 2 | 1 | 1 | 3 | 0.180 | 0.185 | 2 | 2.180 | 2.185 |
| 2 | 0 | 0 | 2 | 0.431 | 0.396 | 2 | 2.431 | 2.396 |
| 3 | 0 | 0 | 3 | 0.500 | 0.447 | 2 | 2.500 | 2.447 |

GA1, GA2 stand for the actual score and GF1, GF2 for the forecast score
D, C and FP are defined in formula 3.
W D L refers to penalty due to category of actual and forecast results.
D1, D2 for L1 and L2 norms respectively; same for FP.

## Discussion & Conclusion

Several choices in the expression of PF can be discussed as there is a large part of arbitrariness in defining a loss function between actual and forecast scores according to the objectives of forecasters.

D defined in (3) is a member of the Minkowski $L_r$-distance $D_r\left(\mathbf{g}^A, \mathbf{g}^F\right) = \left\|\tilde{\mathbf{g}}^A - \tilde{\mathbf{g}}^F\right\|_r / (\left\|\tilde{\mathbf{g}}^A\right\|_r + \left\|\tilde{\mathbf{g}}^F\right\|_r)$ where $\left\|\mathbf{u}\right\|_r = \left(\sum_{i=1}^{2} |u_i|^r\right)^{1/r}$. Another classical choice could be the $L_1$ Manhattan distance although $D_1$ is no longer a metric (Deza & Deza, 2009).

In such a case, $MFP_1$ corresponds to the (unidimensional) Symmetric Mean Absolute Percent Error (SMAPE) used in Business and Econometrics and defined as (Hyndman & Arthanasopoulos, 2013):

$$SMAPE = \frac{1}{n} \sum_{t=1}^{n} \frac{|F_t - A_t|}{|F_t| + |A_t|}$$

There is not much difference if any between L1 and L2 versions of Distance and Penalty for the example of Table 1. It is much a matter of tradition, L1 for instance being popular in Economics while L2 is much used in Genetics. Baker & McHale (2013) reported the (L1) Mean Absolute Deviation (MAD) for predicted points in a comparative study of two models for forecasting exact match scores in American football from data of the National Football League (NFL).

Theoretically $c_0$ must be equal or larger to 1 to ban any possible overlapping. But, football being a low scoring game with tiny probabilities of extreme scores, $c_0$ can be set to a lower value such as $c_0 = \frac{1}{2}$ to shrink penalty values among doublets of W, D, L categories, thus modifying the contributions of the two components C and D of PF.

One might also have envisioned asymmetric penalties regarding W and L to stick close to the point rating of W, D and L i.e., 3, 1, 0 points respectively using for instance the following rating:

$C = 0$ if actual $\mathbf{g}^A$ and forecast $\mathbf{g}^F$ are in the same category (W or D or L),

$C = c_0$ if actual and forecast are in the adjacent categories (D, L) or (L, D),

$C = 2c_0$ if actual and forecast are in the adjacent categories (W, D) or (D, W),

$C = 3c_0$ if actual and forecast are in opposite categories (W, L) or (L, W).

We tried to solve some inconsistencies in evaluating the goodness of score forecasting on the sole basis of usual distance measures by considering not only the score itself, but the ordinal classification of the issue resulting from the score i.e. Win > Draw > Loss in the light of Sutcliffe (1986) differential ordering.

To some extent, it looks like defining a selection index in animal or plant genetics where you must choose an objective among many possible ones (Hazel, 1943). Here, a possibility would be to act as an average gambler who is putting some amount of money on different betting types (1X2, Double chance, Draw no bet, Handicap, Exact score, Total score, etc..). Comparison among exact score forecasting criteria then reduces to select the ones minimizing the expected loss. No doubt that such considerations will stimulate the discussion, especially in the perspective of the 2018 FIFA World Cup in Russia.

**Acknowledgements**

Thanks are expressed to Dr Louis OLLIVIER (Geneticist, Jouy en Josas, France) and Larry SCHAEFFER (University of Guelph, Canada) for their comments and suggestions on an earlier version of the manuscript.